\documentclass[preprint,12pt]{elsarticle}

\usepackage{amsmath,amssymb}
\usepackage[all]{xy}
\usepackage{multirow,booktabs}
\usepackage{etoolbox}
\usepackage{overpic}
\usepackage[colorlinks=true,linkcolor=blue,citecolor=blue,urlcolor=blue,]{hyperref}

\journal{}
\graphicspath{{Figures/}}

\newcommand{\motif}[1]{
	\ifstrequal{#1}{6}{\begin{xy}
			\POS (0,3) *\cir<1.5pt>{} ="a", (-3,-1)*\cir<1.5pt>{} ="b", (3,-1)*\cir<1.5pt>{} ="c"  
			\POS "a" \ar @{->} "b" \POS "a" \ar @{->} "c" \end{xy} }
	
	\ifstrequal{#1}{12}{\begin{xy}
			\POS (0,3) *\cir<1.5pt>{} ="a", (-3,-1)*\cir<1.5pt>{} ="b", (3,-1)*\cir<1.5pt>{} ="c"
			\POS "a" \ar @{<-} "b"  \POS "a" \ar @{->} "c"  \end{xy}  }
	
	\ifstrequal{#1}{14}{\begin{xy}
			\POS (0,3) *\cir<1.5pt>{} ="a", (-3,-1)*\cir<1.5pt>{} ="b", (3,-1)*\cir<1.5pt>{} ="c"
			\POS "a" \ar @{<->} "b"  \POS "a" \ar @{->} "c"  \end{xy}  }
	
	\ifstrequal{#1}{36}{\begin{xy}
			\POS (0,3) *\cir<1.5pt>{} ="a", (-3,-1)*\cir<1.5pt>{} ="b", (3,-1)*\cir<1.5pt>{} ="c"
			\POS "a" \ar @{->} "c"  \POS "b" \ar @{->} "c"  \end{xy}  }
	
	\ifstrequal{#1}{38}{\begin{xy}
			\POS (0,3) *\cir<1.5pt>{} ="a", (-3,-1)*\cir<1.5pt>{} ="b", (3,-1)*\cir<1.5pt>{} ="c"
			\POS "a" \ar @{->} "b"  \POS "a" \ar @{->} "c"  \POS "b" \ar @{->} "c"  \end{xy}  }
	
	\ifstrequal{#1}{46}{\begin{xy}
			\POS (0,3) *\cir<1.5pt>{} ="a", (-3,-1)*\cir<1.5pt>{} ="b", (3,-1)*\cir<1.5pt>{} ="c"
			\POS "a" \ar @{<->} "b"  \POS "a" \ar @{->} "c"  \POS "b" \ar @{->} "c"  \end{xy}  }
	
	\ifstrequal{#1}{74}{\begin{xy}
			\POS (0,3) *\cir<1.5pt>{} ="a", (-3,-1)*\cir<1.5pt>{} ="b", (3,-1)*\cir<1.5pt>{} ="c"
			\POS "a" \ar @{<->} "b"  \POS "a" \ar @{<-} "c"  \end{xy}  }
	
	\ifstrequal{#1}{78}{\begin{xy}
			\POS (0,3) *\cir<1.5pt>{} ="a", (-3,-1)*\cir<1.5pt>{} ="b", (3,-1)*\cir<1.5pt>{} ="c"  
			\POS "a" \ar @{<->} "b" \POS "a" \ar @{<->} "c" \end{xy} }
	
	\ifstrequal{#1}{98}{\begin{xy}
			\POS (0,3) *\cir<1.5pt>{} ="a", (-3,-1)*\cir<1.5pt>{} ="b", (3,-1)*\cir<1.5pt>{} ="c"  
			\POS "a" \ar @{->} "b"  \POS "a" \ar @{<-} "c"  \POS "b" \ar @{->} "c"  \end{xy}  }
	
	\ifstrequal{#1}{102}{\begin{xy}
			\POS (0,3) *\cir<1.5pt>{} ="a", (-3,-1)*\cir<1.5pt>{} ="b", (3,-1)*\cir<1.5pt>{} ="c"  
			\POS "a" \ar @{->} "b"  \POS "a" \ar @{<->} "c"  \POS "b" \ar @{->} "c"  \end{xy}  }
	
	\ifstrequal{#1}{108}{\begin{xy}
			\POS (0,3) *\cir<1.5pt>{} ="a", (-3,-1)*\cir<1.5pt>{} ="b", (3,-1)*\cir<1.5pt>{} ="c"  
			\POS "a" \ar @{<-} "b"  \POS "a" \ar @{<->} "c"  \POS "b" \ar @{->} "c"  \end{xy}  }
	
	\ifstrequal{#1}{110}{\begin{xy}
			\POS (0,3) *\cir<1.5pt>{} ="a", (-3,-1)*\cir<1.5pt>{} ="b", (3,-1)*\cir<1.5pt>{} ="c"  
			\POS "a" \ar @{<->} "b"  \POS "a" \ar @{<->} "c"  \POS "b" \ar @{->} "c"  \end{xy}  }
	
	\ifstrequal{#1}{238}{\begin{xy}
			\POS (0,3) *\cir<1.5pt>{} ="a", (-3,-1)*\cir<1.5pt>{} ="b", (3,-1)*\cir<1.5pt>{} ="c"  
			\POS "a" \ar @{<->} "b"  \POS "a" \ar @{<->} "c"  \POS "b" \ar @{<->} "c"  \end{xy}  }
}

\begin{document}

	\begin{frontmatter}
		
		\title{Motif analysis and passing behavior in football passing networks}

		\author[SS,RCE]{Ming-Xia Li} 
		\author[SS]{Li-Gong Xu\corref{cor1}}
		\ead{xuligong@ecust.edu.cn}
		\author[RCE,SB,SM]{Wei-Xing Zhou\corref{cor1}}
		\ead{wxzhou@ecust.edu.cn}
		
		\cortext[cor1]{Corresponding author}
		\affiliation[SS]{organization={School of Sports Science and Engineering, East China University of Science and Technology},
			city={Shanghai},
			postcode={200237}, 
			country={China}}
		\affiliation[RCE]{organization={Research Center for Econophysics, East China University of Science and Technology},
			city={Shanghai},
			postcode={200237}, 
			country={China}}
		\affiliation[SB]{organization={School of Business, East China University of Science and Technology},
			city={Shanghai},
			postcode={200237}, 
			country={China}}
		\affiliation[SM]{organization={School of Mathematics, East China University of Science and Technology},
			city={Shanghai},
			postcode={200237}, 
			country={China}}
		\begin{abstract}
			The strategic orchestration of football matchplays profoundly influences game outcomes, motivating a surge in research aimed at uncovering tactical nuances through social network analysis. In this paper, we delve into the microscopic intricacies of cooperative player interactions by focusing on triadic motifs within passing networks. Employing a dataset compiled from 3,199 matches across 18 premier football competitions, we identify successful passing activities and construct passing networks for both home and away teams. Our findings highlight a pronounced disparity in passing efficiency, with home teams demonstrating superior performance relative to away teams. Through the identification and analysis of 3-motifs, we find that the motifs with more bidirectional links are more significant. It reveals that footballers exhibit a strong tendency towards backward passes rather than direct forward attacks. Comparing the results of games, we find that some motifs are related to the goal difference. It indicates that direct and effective forward passing significantly amplifies a team's offensive capabilities, whereas an abundance of passbacks portends an elevated risk of offensive futility. These revelations affirm the efficacy of network motif analysis as a potent analytical tool for unveiling the foundational components of passing dynamics among footballers and for decoding the complex tactical behaviors and interaction modalities that underpin team performance.
		\end{abstract}
		
		
		
	\end{frontmatter}

	\section{Introduction}
	\label{S1:Introduction}
	
	
	In the realm of competitive sports, particularly within the context of football, investigating pass behaviors serves as a pivotal avenue for elucidating intricate team dynamics and tactical maneuvers \cite{Duch-Waitzman-Amaral-2010-PLoSOne,Borges-daCosta-RamosSilva-Moura-Serassuelo-Moreira-Praca-Ronque-2023-PerceptMotSkills}. Cho et al. presented a football win-lose prediction system based on passing activities \cite{Cho-Yoon-Lee-2018-EngApplArtifIntell}. Chacoma et al. investigated the intricate choreography of ball possession within the strategic landscape of football \cite{Chacoma-Almeira-Perotti-Billoni-2020-PhysRevE}. Yamamoto and Narizuka investigated the use of Markov-chain models as a tool for understanding the complex dynamics within football games, specifically focusing on the patterns and evolution of ball-passing networks over time \cite{Yamamoto-Narizuka-2018-PhysRevE}. They also  establish a theoretical model to explain the distribution of ball possession times among players and teams \cite{Yamamoto-Uezu-Kagawa-Yamazaki-Narizuka-2024-PhysRevE}, providing a framework for interpreting possession statistics.
	
	Naturally, a passing network is formed by connecting different footballers with passing routes. In sports, many people believed that social networks would provide new sights for sports research \cite{Quatman-Chelladurai-2008-JSportManage,Loughead-Fransen-VanPuyenbroeck-Hoffmann-DeCuyper-Vanbeselaere-Boen-2016-JSportsSci,Mclean-Salmon-Gorman-Dodd-Solomon-2019-SciMedFootball}. Jone et al. investigated the collaborative relationships between non-profit youth sports organizations and found the configuration and structural characteristics of the network \cite{Jones-Edwards-Bocarro-Bunds-Smith-2017-JSportManage}. Bruner et al. explored the social relationship within sport teams by using social network analysis \cite{Bruner-McLaren-Mertens-Steffens-Boen-McKenzie-Haslam-Fransen-2022-PsycholSportExerc}. 
	Using the passing networks, one can reveal the team's passing mode \cite{Yamamoto-Narizuka-2021-PhysRevE}, tactical style \cite{Chacoma-Billoni-Kuperman-2022-PhysRevE}, and offensive intention. Analyzing the passing networks can help the team not only find the links with low passing efficiency but also optimize the team's passing tactics and cooperation methods \cite{Chacoma-Almeira-Perotti-Billoni-2021-PhysRevE}. At the same time, one can also identify the tacit understanding, passing path, and space utilization between players through network analysis, so as to improve the offensive efficiency and scoring ability of the team. Li et al. analyzed the structural attributes of the global football transfer network and found that network attributes may reflect the importance of football clubs in the global transfer market \cite{Li-Xiao-Wang-Zhou-2018-IntJModPhysB,Li-Zhou-Stanley-2019-EPL}. Cocco et al. studied the participation network of football fans and found that the commitment of other fans to the team, members of the supporter group, age, and interaction with other fans in the team environment were associated with higher levels of participation \cite{Cocco-Katz-Hambrick-2021-IntJEnvironResPublicHealth}. By constructing a pass network using a dataset of 3032 passes between teammates in the 2014 FIFA World Cup, Clemente et al. found that the attacking process of the German national football team is based on positional attacks rather than counterattacks, with midfielders being the main players, followed by center backs \cite{Clemente-Silva-Martins-Kalamaras-Mendes-2016-ProcInstMechEngPartP-JSportEngTechnol}. Social network analysis has been an effective way to study the performance of footballers and teams \cite{Li-Xu-Zhou-2023-EPL}.
	
	In social network analysis, motifs can be regarded as the basic modules of network structure \cite{Milo-ShenOrr-Itzkovitz-Kashtan-Chklovskii-Alon-2002-Science,Li-Palchykov-Jiang-Kaski-Kertesz-Micciche-Tumminello-Zhou-Mantegna-2014-NewJPhys}. The importance of motifs is that they may be closely related to the function and dynamic behavior of the network. For example, in biological networks, specific motifs may correspond to signal transduction pathways or regulatory mechanisms in cells \cite{Samaniego-Franco-2018-ACSSynthBiol}. In social networks, motifs may reflect specific interpersonal relationship patterns \cite{Xie-Li-Jiang-Zhou-2014-SciRep}. Motifs usually involve relatively few nodes (such as 3 or 4 nodes). Sinha et al. used network motifs to capture interaction patterns between users in online social networks \cite{Sinha-Bhattacharya-Roy-2022-JSupercomput}. Li et al. believed that network motif was a good tool to investigate the evolution of local relationship patterns \cite{Liu-Zhao-Fu-Cao-2021-IntJProjManag}. By using network motif analysis, Yu et al. studied the influence of network topology on the trading network structure \cite{Yu-Xiong-Xiao-He-Peng-2022-ApplMathComput}. When motifs of higher order (typically of order 4 or 5) are considered, the number of different motifs becomes extremely huge, which makes the definition and identification of motifs more complex. For the directed graph composed of three nodes (called 3-motif), there are 13 different connection modes in theory, but in the actual complex network, not all of these modes appear at the same frequency. Those patterns with significantly high frequencies are the motifs of interest to researchers. In order to identify motifs, researchers usually use algorithms to compare the frequency of motifs in actual networks and a large number of randomized networks \cite{Yu-Feng-Zhang-Bedru-Xu-Xia-2020-ComputSciRev} so as to determine which patterns are significantly non-random.
	
	Football games rely heavily on collective cooperation to achieve both offensive and defensive objectives. The motifs are important to make clear the relationship among footballers from a microscopic perspective. There are some studies investigating the flow motifs in football passing networks \cite{Gyarmati-Hwak-Rodriguez-2014-XXX,Pena-Navarro-2015-XXX,Bekkers-Dabadghao-2019-JSA}. The flow motif focuses on passing events within a fixed time window, which can help people understand the passing behavior characteristics of the team. However, it is not possible to effectively identify individual players' preferences for passing behavior. 
	In this study, we investigated network motifs composed of various footballers. Moreover, by dividing the set of passing networks into home and away categories, we aim to explore how these motifs differ in terms of frequency and structure according to whether a team is playing at home or away, providing deeper insights into the dynamics of professional football games.
	
	\section{Methods}
	
	\subsection{Data description}
	
	We investigate the StatsBomb Open Data. The data we used in this paper contains 3199 football games from 18 different competitions. All pass events in each game and game information can be accessed. Each pass event contains information such as the initiation time of the pass, the passer's name, the name of the team the passer belongs to, the receiver's name, the name of the team the receiver belongs to, the location of the passer, and the location of the receiver. The information for each game contains the names of the home and away teams and the result of the game.

	\subsection{Construction of passing network}
	
\begin{figure}[htp]
		\centering
		\includegraphics[width=0.8\linewidth,height=0.6\linewidth]{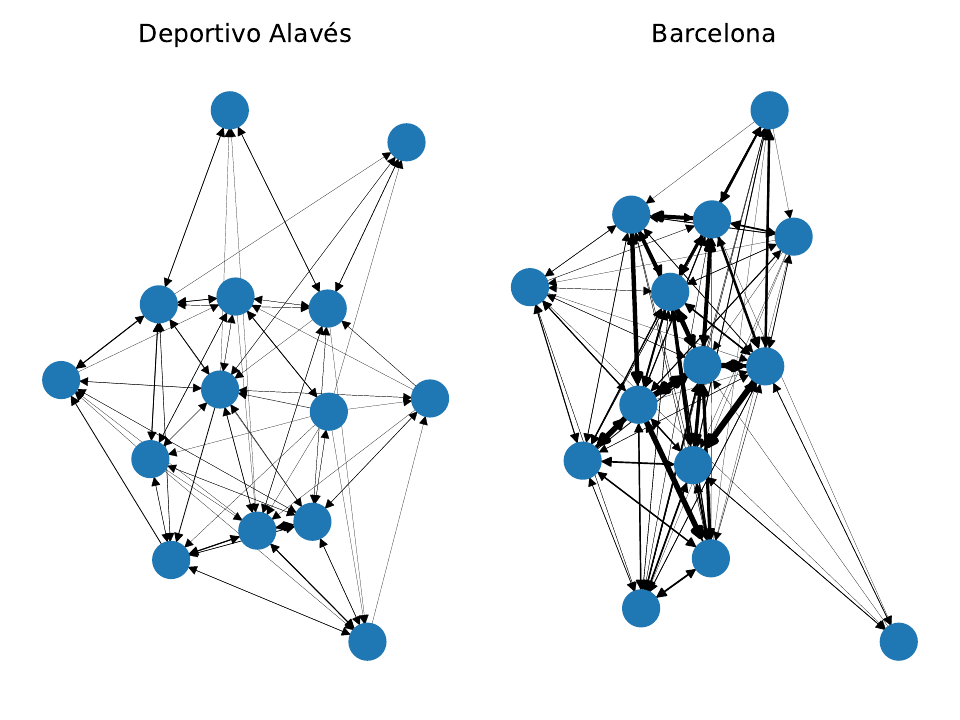}
		\caption{Passing networks of two teams, Deportivo Alaves (home team) and Barcelona (away team), in one game. Each node refers to a specific footballer. The direction of each link refers to the direction of passes between two footballers. The width of each link corresponds to the number of passes.}
		\label{fig:passnet}
	\end{figure}

	A network is composed of a couple of nodes connected to each other.
	In our passing network, nodes are the specific footballers. When a footballer makes a pass to another footballer, a directed link connects the passer to the receiver. Here, we only consider the success passes during each game. It indicates that a pair of passing and receiving footballers belong to the same team, and one football game corresponds to two passing networks. One is the home passing network for the home team, and the other is the away passing network for the away team. Fig.~\ref{fig:passnet} shows the passing networks for a home team and an away team in a football game. Due to the presence of substitute players, there may be more than 11 players in a passing network.
	
	\subsection{Network motif analysis}
	
	A network motif can be considered a local connection pattern that frequently occurs in a network. In this paper, we investigate the motifs composed of three nodes connected with directed links, called 3-motif. In this case, a directed network has no more than 13 types of 3-motifs, which are listed in Table~\ref{table:3motif}. We use the method of Milo et al. to label the id of these 3-motifs \cite{Milo-ShenOrr-Itzkovitz-Kashtan-Chklovskii-Alon-2002-Science}. 
	
	\begin{table}[htp]
		\centering
		\caption{List of 3-motifs. The second column is the adjacency matrix of each network motif.}
		\label{table:3motif}
		
		\begin{tabular}{ccc}
			\hline\hline
			motif&  adjacency matrix & motif id\\
			\hline
			\motif{6} & 000000110 & 6\\
			\motif{12} & 000001100 & 12\\
			\motif{14} & 000001110 & 14\\
			\motif{36} & 000100100 & 36\\
			\motif{38} & 000100110 & 38\\
			\motif{46} & 000101110 & 46\\
			\motif{74} & 001001010 & 74\\
			\motif{78} & 001001110 & 78\\
			\motif{98} & 001100010 & 98\\
			\motif{102} & 001100110 & 102\\
			\motif{108} & 001101100 & 108\\
			\motif{110} & 001101110 & 110\\
			\motif{238} & 011101110 & 238\\
			\hline\hline
		\end{tabular}
	\end{table}
	
	For each football game, we have two passing networks. One is the passing network for the home team. The other is the passing network for the away team. For each passing network, we use the ``gtrieScanner'' provided by Pedro Ribeiro \cite{Ribeiro-Silva-2014-DataMinKnowlDiscov} to detect the 3-motifs. Then, we have the number of each 3-motif for each passing network.
	
	Since the number of motifs is related not only to the local relationships of nodes but also to the topological structure of the passing network, we investigate the significance of 3-motifs in a network by comparing 100 random networks. Random networks are obtained by shuffling the links of a network, which can maintain the degree distribution of the network.
	
	\section{Results}
	
	\subsection{Statistic analysis of football games}
	
	For any competition, the final result is definitely the most important. Among these 3199 football games, 1452 are won by home teams, accounting for 45.4\%. 1018 games are won by away teams, accounting for 31.8\%. The rest of the games are draws, accounting for 22.8\%. It indicates that home teams have a higher winning rate than away teams. A common reason is that home teams have more favorable conditions than away teams.
	
	\begin{figure}[htp]
		\centering
		\includegraphics[width=0.7\linewidth]{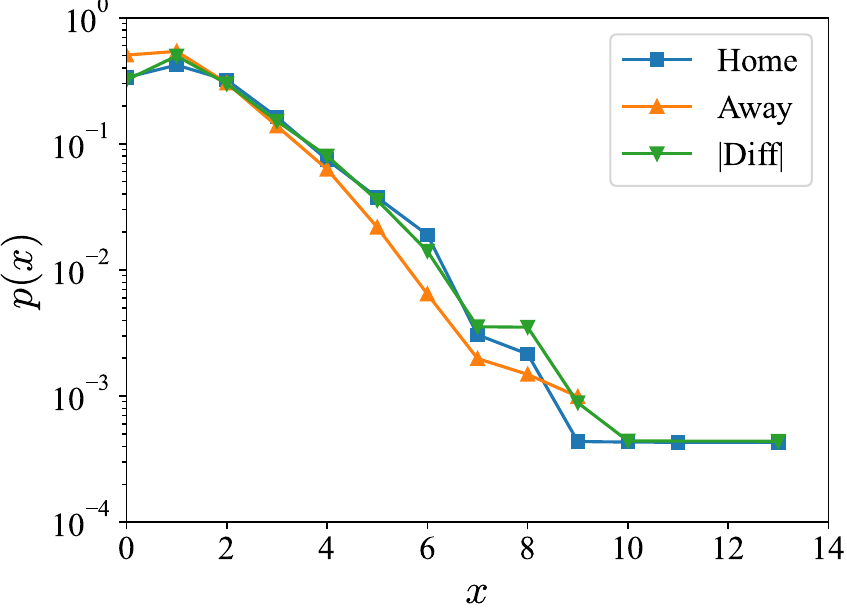}
		\caption{Probability density distributions of the number of goals. Home refers to the number of goals scored by home teams. Away refers to the number of goals scored by away teams. $|\rm{Diff}|$ refers to the absolute value of the goal differential.}
		\label{fig:pdf:score}
	\end{figure}
	
	\begin{table}[htp]
		\centering
		\caption{Statistical results of the number of goals.}
		\label{table:stat:score}
		\begin{tabular}{cccccccccc}
			\hline\hline
			& mean & std & median & min & max & kurtosis & skewness\\
			\hline
			Home & 1.600 & 1.460 & 1.000 & 0.000 & 13.000 & 3.031 & 1.302\\
			Away & 1.260 & 1.266 & 1.000 & 0.000 & 9.000 & 2.525 & 1.319\\
			Diff & 0.340 & 2.092 & 0.000 & -9.000 & 13.000 & 1.392 & 0.083\\
			\hline\hline
		\end{tabular}
	\end{table}
	
	Furthermore, we investigate the details of the number of goals for home and away teams. We also calculate the difference in the number of goals between home and away teams, regarded as the goal difference for the home team. In Fig.~\ref{fig:pdf:score}, the number of goals is approximated by exponential distributions.
	The statistical results of the number of goals are shown in Table~\ref{table:stat:score}. The average number of goals scored by home and away teams is $1.600\pm1.460$ and $1.260\pm 1.266$, respectively. It indicates that most teams can only score one or two goals in a game. Compared with away teams, home teams get more goals on average.

	In football games, passing is a fundamental part of football activities. Lots of studies focused on the offense process in the games \cite{Duch-Waitzman-Amaral-2010-PLoSOne,Buldu-Busquets-Echegoyen-Seirullo-2019-SciRep,Praca-Rochael-Francklin-daSilva-deAndrade-2021-ProcInstMechEngPartP-JSportEngTechnol,Borges-daCosta-RamosSilva-Moura-Serassuelo-Moreira-Praca-Ronque-2023-PerceptMotSkills}. It is started from the moment a player gains possession of the ball and ends when they lose possession or score a goal, called the ``unit of attack''. The number of units of attacks is the number of offensive plays in the game, reflecting the enthusiasm of the team's offense. The length of the unit of attack represents the ball control of a team. Since the football network in this article includes only successful passes, the passing network can also be viewed as consisting of these unit attacks.
	
	\begin{figure}[htp]
		\centering
		\includegraphics[width=0.49\linewidth]{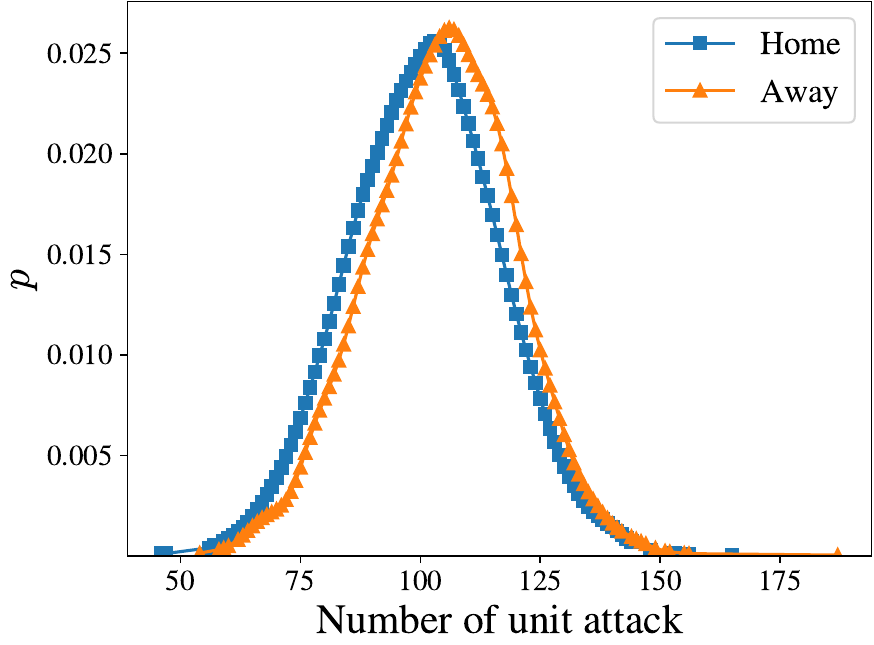}
		\includegraphics[width=0.49\linewidth]{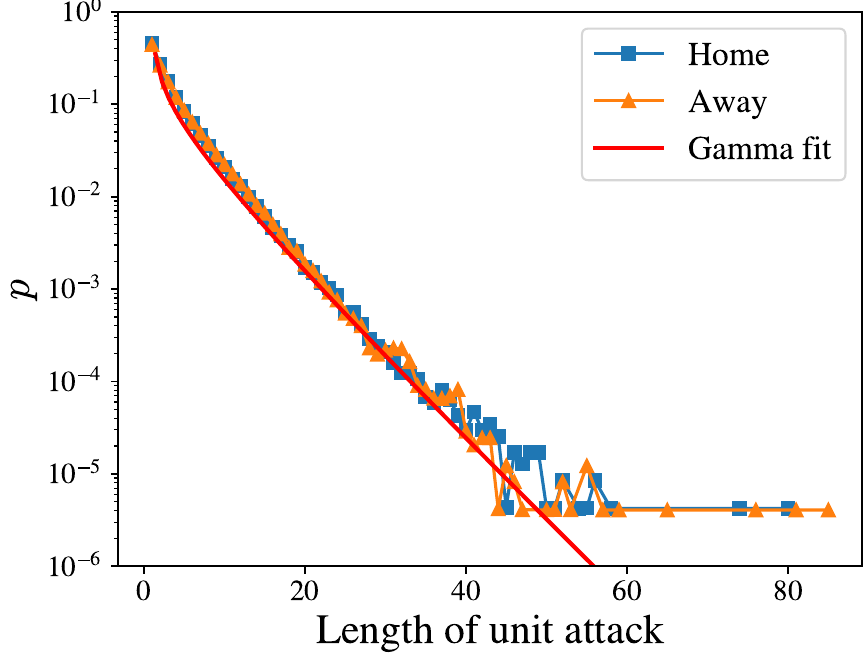}
		\caption{Probability density distributions of the number of units of attacks in each game (left) and the length of each unit of attack (right). The number of units of attacks is the number of offense processes in a game. Length of unit attack is the number of passes in one offense process. The distributions of the number of units of attacks in each game are close to a normal fit. The distributions of length of each unit of attacks are close to a gamma fit with $\alpha = 0.515$ and $\beta = 0.191$.}
		\label{fig:pdf:UA}
	\end{figure}

	As shown in Fig.~\ref{fig:pdf:UA}, the average number of units of attacks in home and away networks is $100.902\pm15.638$ and $104.379\pm15.254$, respectively. Home teams have made fewer units of attacks than away teams. The right plot in Fig.~\ref{fig:pdf:UA} shows the distributions of length for each unit of attacks. The distributions are observed to be fitted as a gamma curve $p(x) = \frac{\beta^\alpha}{\Gamma(\alpha)}x^{\alpha-1}e^{-\beta x}$, where the shape parameter $\alpha$ is $0.515$ and the inverse scale parameter $\beta$ is $0.191$. 

	\subsection{Statistic analysis of passing networks}

\begin{figure}[htp]
		\centering
		\includegraphics[width=0.7\linewidth]{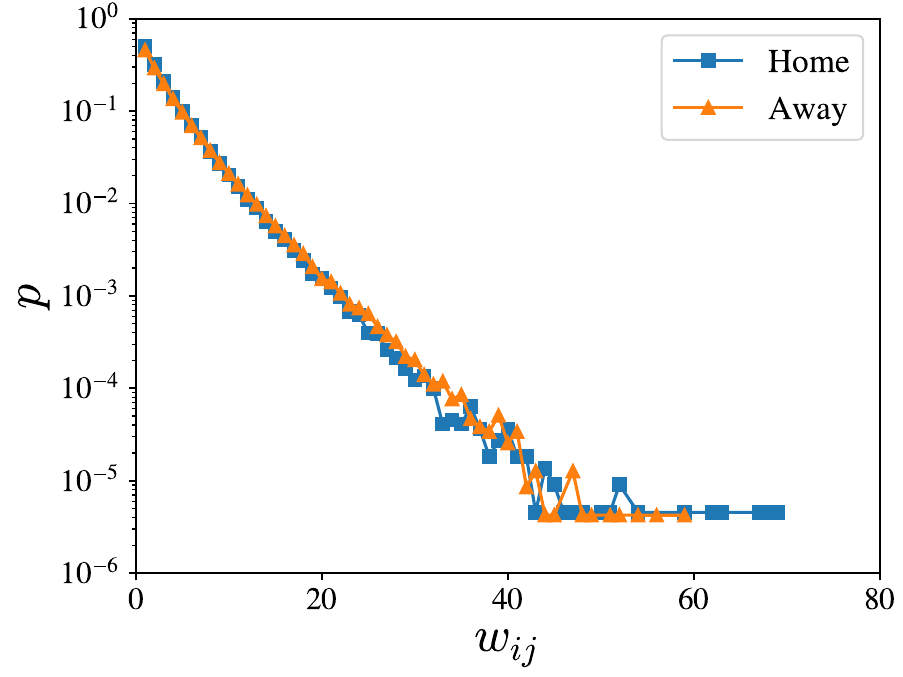}
		\caption{PDF of link weights in all passing network. Link weight $w_{ij}$ is the number of passes from footballer $i$ to $j$ in one game.}
		\label{fig:pdf:weight}
	\end{figure}

	In a passing network, we call the number of passes from footballer $i$ to footballer $j$ the weight of link $w_{ij}$. The stronger the edge weight, the closer the passing relationship between the two footballers. According to the distribution of link weights (in Fig.~\ref{fig:pdf:weight}), one can see that the probability of the occurrence of links decreases with a nearly exponential trend as the weight increases. More than half of the links have weights less than 2 (weak ties). Only about 5\% of links have weights greater than 10. It implies that most passing relationships may not be formed due to passing relationships between footballers but rather are randomly generated because the number of nodes in a passing network is limited and the number of passes is huge. It will lead to a highly connected passing network. 
	
	\begin{figure}[htp]
		\centering
		\includegraphics[width=0.32\textwidth,height=0.25\textwidth]{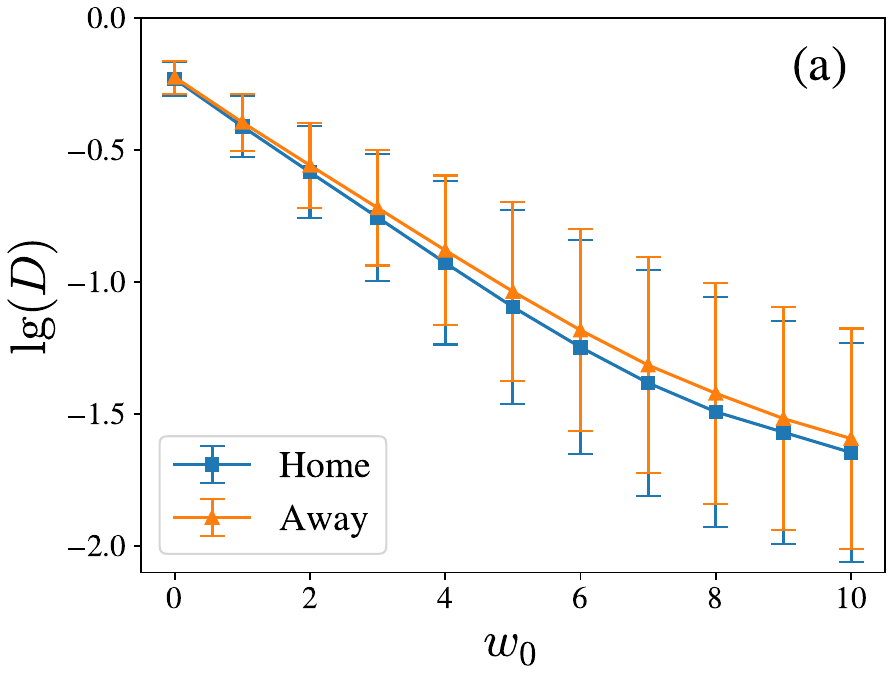}
		\includegraphics[width=0.32\textwidth,height=0.25\textwidth]{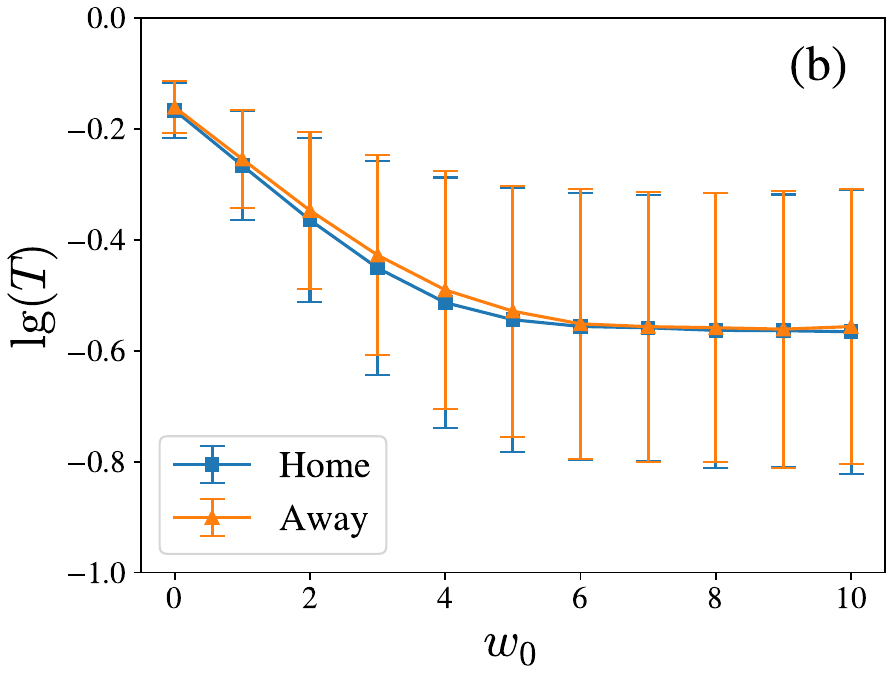}
		\includegraphics[width=0.32\textwidth,height=0.25\textwidth]{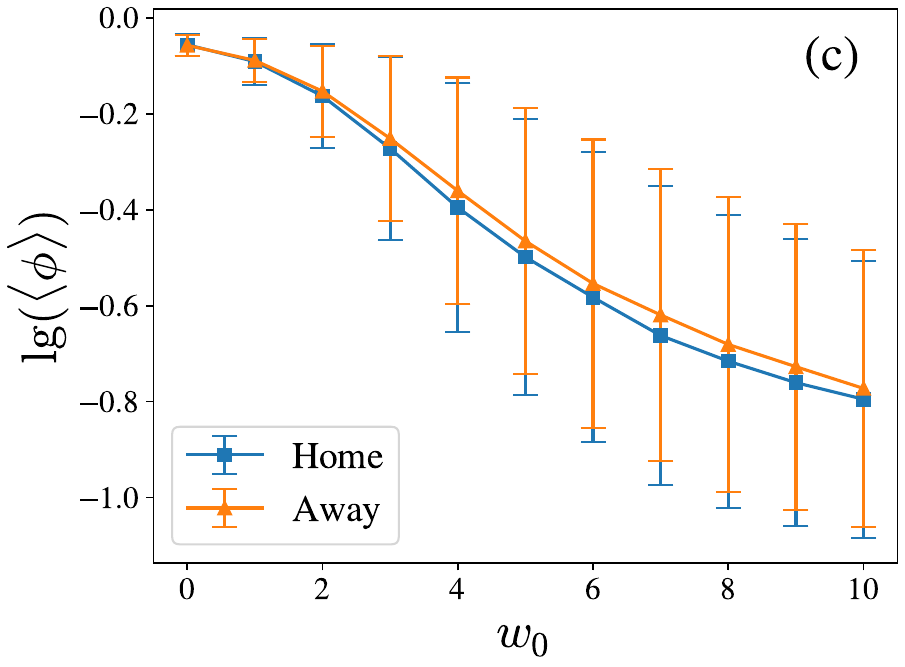}\\
		\includegraphics[width=0.32\textwidth,height=0.25\textwidth]{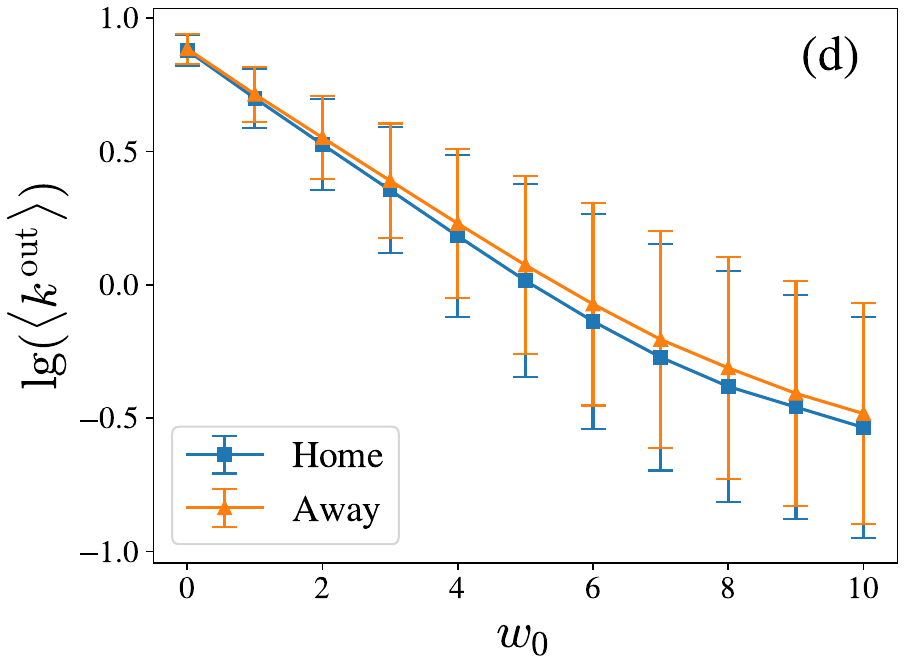}
		\includegraphics[width=0.32\textwidth,height=0.25\textwidth]{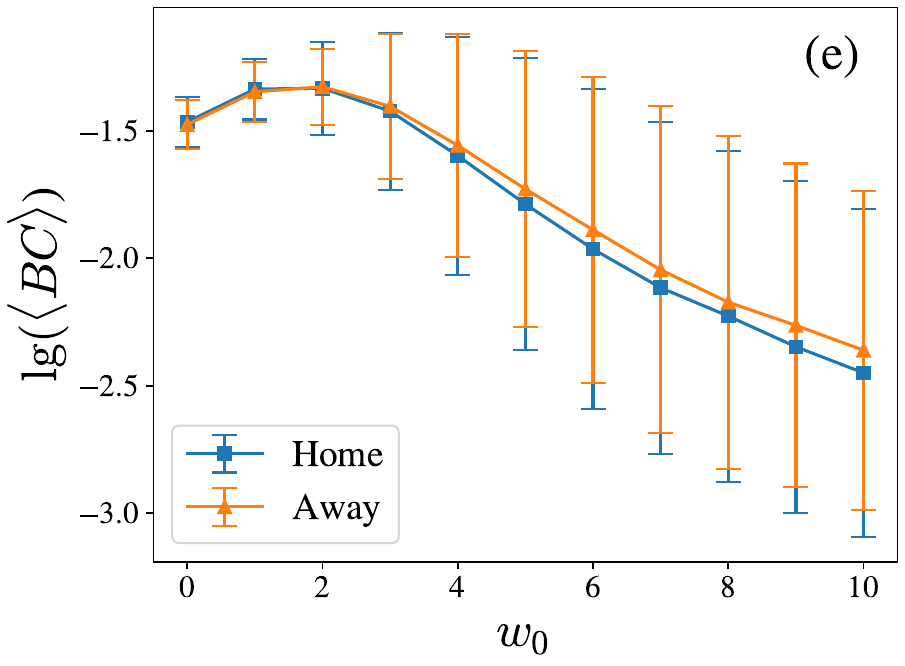}
		\includegraphics[width=0.32\textwidth,height=0.25\textwidth]{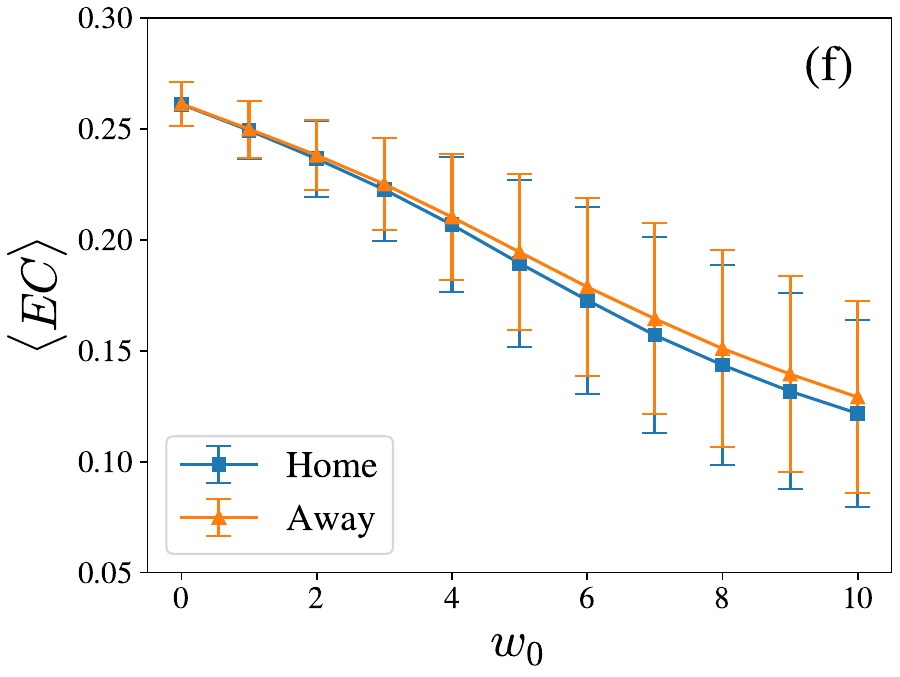}
		\caption{Errorbar plots for different network measures of all passing networks. $w_0$ is the threshold of link weight. The network measures from plot (a) to plot (e) are the base-10 logarithms of network density $D$, transitivity $T$, average diversity $\langle \phi \rangle$, average out-degree centrality $\langle k^{\rm{out}}$ and average betweenness centrality $\langle BC \rangle$. The plot (f) refers to the average eigenvector centrality $\langle EC \rangle$. Each marker refers to the average value of the corresponding network measure in the passing networks, in which the weights of links are larger than $w_0$. The length of the error line on each marker refers to the corresponding standard deviation.}
		\label{fig:netvar}
	\end{figure}

	Similar to social network analysis, big data usually contain erroneous information \cite{Li-Palchykov-Jiang-Kaski-Kertesz-Micciche-Tumminello-Zhou-Mantegna-2014-NewJPhys}. In passing processes, a wrong pass is a possible source of false links in the construction of passing networks. Unless data are cleaned, spurious links could be misinterpreted as real passing relationships. 
	To identify the real cooperative relationship between players, it is necessary to construct the passing network by filtering the links of different weights. We construct different passing networks with different thresholds of link weight $w_0$. In those passing networks, there is a passing relationship between two footballers only when the number of passes between them exceeds $w_0$ (i.e., the weight of each link is greater than $w_0$). Fig.~\ref{fig:netvar} and Table~\ref{table:stat:netvar} show the network characteristics of passing networks with different thresholds of link weight $w_0$.

	\begin{table}[htp]
		\centering
		\caption{Average values of network metrics of the passing networks with different thresholds of link weight $w_0$ for both home and away teams. The network metrics in the table are the number of links $L$, network density $D$, network transitivity $T$, average value of node diversity $\langle \phi \rangle$, average value of node outdegree centrality $\langle k^{\rm{out}} \rangle$, average value of node betweenness centrality $\langle BC \rangle$ and average value of node eigenvector centrality $EC$.}
		\label{table:stat:netvar}
		\begin{tabular}{ccccccccccc}
			\hline\hline
			& ishome & $w_0=0$ & $w_0=2$ & $w_0=10$ \\
			\hline
			$L$ & Home & 339439 & 159249 & 14381 & \\
			& Away & 344074 & 167236 & 17159 & \\
			$D$ & Home & 0.594 $\pm$ 0.084 & 0.280 $\pm$ 0.097 & 0.025 $\pm$ 0.032 & \\
			& Away & 0.603 $\pm$ 0.082 & 0.294 $\pm$ 0.095 & 0.030 $\pm$ 0.034 & \\
			$T$ & Home & 0.686 $\pm$ 0.074 & 0.453 $\pm$ 0.130 & 0.063 $\pm$ 0.147 & \\
			& Away & 0.695 $\pm$ 0.071 & 0.469 $\pm$ 0.127 & 0.077 $\pm$ 0.160 & \\
			$\langle \phi \rangle$ & Home & 0.881 $\pm$ 0.045 & 0.705 $\pm$ 0.137 & 0.080 $\pm$ 0.127 & \\
			& Away & 0.879 $\pm$ 0.043 & 0.718 $\pm$ 0.128 & 0.100 $\pm$ 0.137 & \\
			$\langle k_{\rm{out}} \rangle$ & Home & 7.633 $\pm$ 0.961 & 3.590 $\pm$ 1.185 & 0.324 $\pm$ 0.403 & \\
			& Away & 7.744 $\pm$ 0.953 & 3.772 $\pm$ 1.164 & 0.387 $\pm$ 0.428 & \\
			$\langle BC \rangle$ & Home & 0.035 $\pm$ 0.008 & 0.050 $\pm$ 0.017 & 0.004 $\pm$ 0.010 & \\
			& Away & 0.034 $\pm$ 0.008 & 0.050 $\pm$ 0.016 & 0.005 $\pm$ 0.010 & \\
			$\langle EC \rangle$ & Home & 0.261 $\pm$ 0.010 & 0.237 $\pm$ 0.017 & 0.122 $\pm$ 0.042 & \\
			& Away & 0.261 $\pm$ 0.010 & 0.238 $\pm$ 0.016 & 0.129 $\pm$ 0.043 & \\
			\hline\hline
		\end{tabular}
	\end{table}
	
	When investigating the characteristics of a football network, people usually use some network measures to observe it \cite{Clemente-Martins-Kalamaras-Wong-Mendes-2015-IJPAS,Goncalves-Coutinho-Santos-LagoPenas-Jimenez-Sampaio-2017-PLoSOne,McLean-Salmon-Gorman-Dodd-Solomon-2021-JHumKinet}. Football games are regarded as a multiplayer sport. Obviously, the cooperation of different players usually has a great impact on the results of football games. We use network density and network transitivity to show the connectivity among footballers in passing networks. In a directed network with $N$ nodes and $L$ links, network density can be defined as the ratio of the number of empirical links to the maximum possible number of links.
	\begin{equation}
		D = \frac{L}{N(N-1)}.
	\end{equation}
	If $D=1$, the passing network is fully connected. All footballers had made passes to all other footballers. If $D=0$, all nodes in the network are separated. No pass occurs among the footballers. In Fig.~\ref{fig:netvar}(a), one can observe that the network density changed with the increase of the link weight threshold $w_0$. Network density for all original passing networks ($w_0=0$) is $0.598$ on average, which is much higher than that in social networks \cite{Li-Jiang-Xie-Micciche-Tumminello-Zhou-Mantegna-2014-SciRep}. When half of the links are removed from the passing networks ($w_0 = 2$), the average network density reduces to $0.3$. It indicates that a footballer would pass balls more than twice with three teammates on average. When $w_0= 10$, the network density is reduced to $0.03$. Only about three pairs of footballers pass the ball more than 10 times. An approximate exponential decreasing trend of the average network density can be found with the increase of $w_0$, which is similar to the distribution of link weight.
	
	Fig.~\ref{fig:netvar}(b) is the correlation between network transitivity $T$ and $w_0$. Network transitivity, also called the global clustering coefficient, is the fraction of all possible triangles presented in a passing network \cite{Sousa-Bredt-Greco-Clemente-Teoldo-Praca-2019-IntJPerformAnalSport}.
	\begin{equation}
		T=\frac{\# \rm{triangles}}{\#\rm{triads}},
	\end{equation}
	where $\#$triangles and $\#$triads are the numbers of triangles and triads. If $T = 1$, all passes between two players are direct passes. If $T=0$, most passes are indirect passes. It means that multiple passes are required to complete a pass from one footballer to another. When $w_0 = 0$, network transitivities in original passing networks are very high, which are close to 0.69 for both home and away teams. When $w_0 = 2$, network transitivities decrease to 0.45 for the home team and 0.47 for the away team.
	
	Motivated by social diversity \cite{Eagle-Macy-Claxton-2010-Science,Jiang-Xie-Li-Podobnik-Zhou-Stanley-2013-ProcNatlAcadSciUSA}, we use passing diversity to quantify how the passers split the number of passes to the receivers:
	\begin{equation}
		\phi_i = \frac{-\sum\limits_{i=1}^{k_i}p_{ij}\log(p_{ij})}{\log(k_i)},
	\end{equation}
	where $k_i$ is the number of receivers of the passes from the passer $i$. $p_{ij}$ is the probability that passer $i$ made a pass to receiver $j$. It is defined as $p_{ij} = w_{ij}/w_i=w_{ij}/\sum_j w_{ij}$, where $w_{ij}$ is the number of passes from passer $i$ to receiver $j$ and $w_i$ is the number of total passes made by passer $i$. A higher $\phi_i$ value indicates that the passer's passes are split more evenly among his receivers, and a smaller $\phi_i$ value implies that most of the passes are passed to only one of his receivers. $\phi = \frac{1}{N}\sum_i \phi_i$ is the average of nodes' diversity values in a network. In Table~\ref{table:stat:netvar}, all teams have a high diversity on average, which is close to $0.88$. It indicates that footballers pass evenly to other players throughout the entire game. Furthermore, from the comparison between the winning and losing teams, it can be seen that footballers in winning teams pass more evenly to other players, while footballers in losing teams pass relatively more concentrated on other players.
	
	Besides the network measures above, we also investigate the node's centrality measures, such as node degree \cite{Yamamoto-Yokoyama-2011-PLoSOne}, betweenness centrality \cite{Korte-Link-Groll-Lames-2019-FrontPsychol}, and eigenvector centrality \cite{HerreraDiestra-Echegoyen-Martinez-Garrido-Busquets-Io-Buldu-2020-ChaosSolitonsFractals}. Node's centrality measures are used to characterize the importance of a node in the network. A node's outdegree in passing networks, $k_i^{\rm{out}}$, is the number of receivers caught the balls from passer $i$ in one game. Generally, a footballer with a high outdegree is the one who will make passes to the majority of other players. This kind of footballer tends to be a playmaker. In table.~\ref{table:stat:netvar}, the average outdegree is nearly $7.7$. It suggests that most footballers had passed with more than seven players. In Fig.~\ref{fig:netvar}(d), the trends of $\langle k^{\rm{out}} \rangle$ are similar to the trends of network density $D$. It is due to the definition of two measures, which are strongly correlated to the number of links. When $w_0$ is 2, the average outdegree is close to 5.24. It suggests that a footballer makes at least two passes to $5.24$ receivers on average. 
	
	Betweenness centrality, denoted by $BC$, is used to quantify the extent to which a node lies on paths between other nodes in a network. For a node in the passing network, its betweenness centrality is calculated as the sum of the fraction of all-pairs shortest paths that pass through that node. Footballers with high betweenness centrality act as crucial intermediaries in attack paths. By closely controlling the important intermediaries in the attack path, footballers can effectively defend the attack. Different from the other network measures, a growth of the average betweenness centrality $\langle BC \rangle$ can be observed in Fig.~\ref{fig:netvar}(e) when $w_0<2$. When $w_0 >2$, $\langle BC \rangle$ changed into a downward trend. It suggests that there were a lot of weak links that do not reflect the real cooperation between footballers. By removing the weak links, the crucial intermediaries can be seen more clearly. 
	
	Eigenvector centrality, denoted by $EC$, is a network measure that evaluates the importance of a node in a network based on the principle that connections to high-scoring nodes contribute more to the score of a node than equal connections to low-scoring nodes. In a passing network, the high eigenvector centrality of a footballer usually comes from his passes with other important footballers. Even though their number of direct passes may not be the largest, the passes they participate in tend to have higher tactical value, connecting the key positions or players in the team. In Fig.~\ref{fig:netvar}(f), the average eigenvector centrality decreases with the increase of $w_0$. It suggests that the weak links would not change the trend of eigenvector centrality.
	
	By comparing the network measures between home teams and away teams, we find that network measures for home teams decreased faster than those for away teams. It suggests that footballers from away teams worked in closer collaboration than from home teams. Comparing to the results in Table~\ref{table:stat:score} that home teams have a higher win rate, we can observe that cooperation between footballers in the winning team is not very close. Teams that gain an advantage usually do not play with all their strength. And backward teams usually make more efforts to change the situation.

	\subsection{Network motifs in passing network}
	
	Network variables can provide a good observation and comparison of the overall characteristics of the network, but they cannot fully represent the microstructural characteristics of the network. Network motifs are a type of network primitive with local connectivity characteristics, each of which has its own specific meaning. By analyzing the characteristics of network motifs, we can gain a deeper understanding of the connectivity characteristics between nodes in the network.
	
	\begin{figure}[htp]
		\centering
		\includegraphics[width=0.7\linewidth]{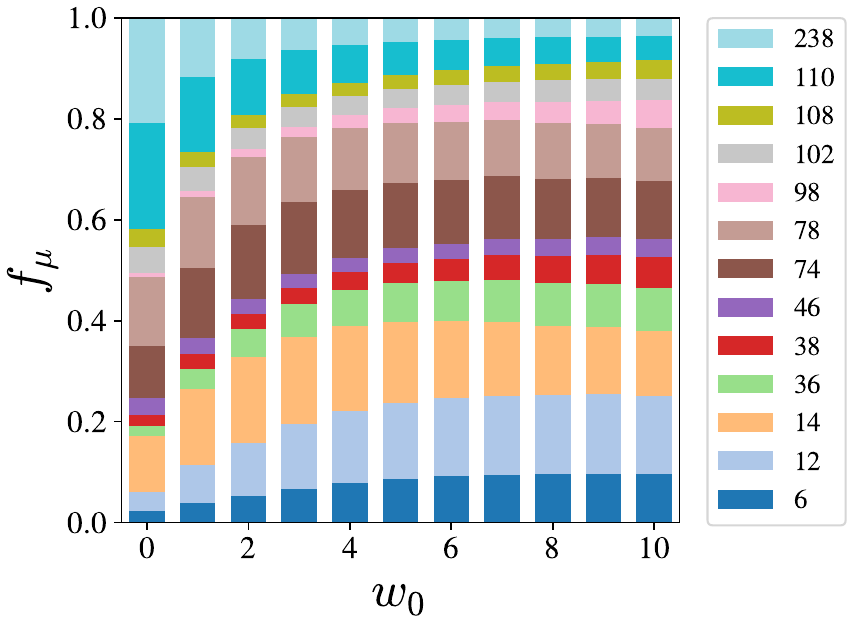}
		\caption{Fraction of average number of 3-motifs $f_{\mu}$ in all networks as a function of the threshold of weight $w_0$.}
		\label{fig:motif:orig:w0}
	\end{figure}
	
	Because of the high density of the original passing network, most nodes are connected to each other. When $w_0 = 0$, the 3-motifs with the largest numbers are coded in 110 and 238. In Fig.~\ref{fig:motif:orig:w0}, the sum of the fractions of motif 238 and 110 is nearly 40\%. With the growth of $w_0$, links with small weights are removed from the passing network. The number of 3-motifs with more links (e.g., motif 110, 238) drops rapidly, while the number of 3-motifs with fewer links (e.g., motif 6, 12, 36) increases. By filtering out a large number of weak connections in the network, strong cooperative relationships among footballers can be identified. This is essential for analyzing team tactical strategies. In the following part, we mainly focus on the passing networks with a link weight threshold $w_0 = 2$.

	\begin{table}[htp]
		\centering
		\caption{Network motifs in passing networks ($w_0=2$). All numbers are the average numbers of different types of passing networks. $\mu$ and $\sigma$ are the average number and standard deviation of 3-motifs in original passing networks. $\mu_{\rm{rnd}}$ and $\sigma_{\rm{rnd}}$ are the average value and standard deviation of random networks. $z$ is the $z$-score value, calculated by $z=(\mu-\mu_{rnd})/\sigma_{rnd}$. The 3-motifs are sorted by the $z$-scores of home networks.}
		\label{table:motif}
			\begin{tabular}{cccccccccccccc}
				\hline\hline
				id & ishome & $\mu$ & $\sigma$ & $\mu_{\rm{rnd}}$ & $\sigma_{\rm{rnd}}$ & $
				z$\\
				\hline
				38 & Home & 3.223 & 2.570 & 6.425 & 2.239 & -1.431\\
				\motif{38} & Away & 3.066 & 2.388 & 6.363 & 2.277 & -1.448\\
				98 & Home & 1.665 & 1.245 & 2.878 & 0.868 & -1.398\\
				\motif{98} & Away & 1.456 & 0.965 & 2.782 & 0.869 & -1.526\\
				102 & Home & 4.558 & 2.334 & 7.677 & 2.260 & -1.380\\
				\motif{102} & Away & 4.478 & 2.250 & 7.943 & 2.170 & -1.597\\
				12 & Home & 11.359 & 7.701 & 18.314 & 8.213 & -0.847\\
				\motif{12} & Away & 10.561 & 7.206 & 17.302 & 7.901 & -0.853\\
				6 & Home & 5.813 & 4.403 & 8.488 & 4.521 & -0.592\\
				\motif{6} & Away & 5.403 & 4.074 & 8.098 & 4.192 & -0.643\\
				108 & Home & 2.666 & 1.638 & 3.293 & 1.143 & -0.549\\
				\motif{108} & Away & 2.625 & 1.589 & 3.354 & 1.139 & -0.640\\
				46 & Home & 3.124 & 1.852 & 3.647 & 1.295 & -0.404\\
				\motif{46} & Away & 3.196 & 1.832 & 3.867 & 1.358 & -0.494\\
				36 & Home & 6.142 & 5.514 & 8.086 & 5.558 & -0.350\\
				\motif{36} & Away & 5.497 & 4.524 & 7.339 & 4.467 & -0.412\\
				110 & Home & 11.471 & 5.085 & 10.371 & 6.135 & 0.179\\
				\motif{110} & Away & 12.025 & 4.956 & 11.133 & 6.103 & 0.146\\
				74 & Home & 15.426 & 5.255 & 12.676 & 3.177 & 0.865\\
				\motif{74} & Away & 15.157 & 5.094 & 12.594 & 3.140 & 0.816\\
				14 & Home & 17.827 & 5.651 & 14.046 & 3.532 & 1.070\\
				\motif{14} & Away & 17.943 & 5.530 & 14.188 & 3.420 & 1.098\\
				238 & Home & 8.096 & 5.980 & 3.922 & 3.898 & 1.071\\
				\motif{238} & Away & 8.718 & 6.063 & 4.286 & 3.925 & 1.129\\
				78 & Home & 14.001 & 7.081 & 6.296 & 2.798 & 2.754\\
				\motif{78} & Away & 14.534 & 7.142 & 6.536 & 2.745 & 2.913\\
				\hline\hline
			\end{tabular}
	\end{table}
	
	In directed networks, 3-motif, a triad with three nodes, is the simplest network motif. It has a total of 13 types, as shown in Table~\ref{table:motif}. Different motifs correspond to different passing activities. e.g., the 3-motif encoded as 6 is more like the starting point of a different attack path. The average number of each motif is denoted by $\mu$. $\sigma$ is the corresponding standard deviation in all home or away networks. 
	
	Because the number of motifs in a network is affected by the network topology, a network with high density will have more 3-motifs with a larger number of links. To compare the 3-motifs in different passing networks, we construct 100 random networks for each passing network by keeping the degrees of all nodes. We identify the 3-motifs in all the random networks. The average number of each 3-motif in all random networks is denoted by $\mu_{\rm{rnd}}$. $\sigma_{\rm{rnd}}$ is the corresponding deviation in all random home or away networks.
	
	For each 3-motif, we evaluate a $z$-score defined as $z=(\mu-\mu_{\rm{rnd}})/\sigma_{\rm{rnd}}$. This variable is used as a statistical measure to assess the significance of observed 3-motifs compared to what would be expected by chance. In Table~\ref{table:3motif}, the 3-motifs are ordered by the $z$-scores of home networks. We find that the 3-motif encoded as 78 had a $z$-score over 1.96 (significance level 0.05). In this 3-motif, the footballer in center is the intermediary of the other two footballers' passing, but there is no direct passing between them. This may reflect the role of a core or transitional player in an organization. He or she receives the ball and assigns it to different teammates, but the two teammates do not directly exchange ball rights.
	
	For other motifs, we find that most of the 3-motifs with bidirectional links have positive $z$-scores. The 3-motifs with no bidirectional links have negative $z$-scores. The rest of the 3-motifs with negative $z$-scores have fewer bidirectional links than unidirectional links. It indicates that the $z$-score increases with the increase in the proportion of bidirectional links in a 3-motif. The bidirectional passes between footballers are more than fixed single-directional passes.
	
	Compared between two teams, most of the 3-motifs (except 3-motifs encoded as 14, 78, 238) for home teams have a larger $z$-score than the 3-motifs for away teams. The $z$-score of 3-motifs with a high ratio of bidirectional links for away teams is larger than that for home teams. It suggests that away teams make more passbacks than home teams.

	\subsection{Correlation between game results and network motifs}
	
	As a competitive sports game, people are most concerned about the final result of the game. Many studies believe that the network measures are related to the final result of the football game \cite{Buldu-Busquets-Echegoyen-Seirullo-2019-SciRep,Aquino-Carling-Vieira-Martins-Jabor-Machado-Santiago-Garganta-Puggina-2020-JStrengthCondRes}. Here, we investigate the correlation between the structural characteristics of the passing network and the game results from a microscopic perspective of the network for all 3199 football games. We construct a multiple linear regression model between the game results and the number of network motifs.
	\begin{equation}
		P=\beta_0+\sum_{i=1}^{26}{\beta_i n_i},
	\end{equation}
	where $P = P_{\rm{home}}-P_{\rm{away}}$ is the goal differential for the home team. $n_i$ is the number of one 3-motif in home or away passing networks for the corresponding game. We use the Ordinary Least Square (OLS) method to obtain the regression coefficient.

	\begin{table}[htp]
		\centering
		\caption{OLS regression results between the goal difference and the number of motifs. The observed value of the F statistic for the OLS model is 29.09 and the $p$-value is 0. $\beta$ is the regression coefficient value of each 3-motif. Std. Error is the standard error. $t$ and $p$ are the T test value and $p$-value for all the 3-motifs.}
		\label{table:motif:ols:freq}
		\begin{tabular}{ccccccccc}
			\hline\hline
			id & ishome &   $\beta$ &  Std. Error &      $t$ &  $p$ \\
			\hline
			const &  & -0.432 & 0.189 & -2.286 & \textbf{0.022}\\
			6 & Home & 0.004 & 0.013 & 0.290 & 0.772\\
			\motif{6} & Away & -0.012 & 0.013 & -0.951 & 0.341\\
			12 & Home & 0.039 & 0.009 & 4.427 & \textbf{0.000}\\
			\motif{12} & Away & -0.023 & 0.009 & -2.506 & \textbf{0.012}\\
			14 & Home & 0.021 & 0.005 & 3.903 & \textbf{0.000}\\
			\motif{14} & Away & -0.008 & 0.005 & -1.503 & 0.133\\
			36 & Home & -0.012 & 0.012 & -0.980 & 0.327\\
			\motif{36} & Away & 0.003 & 0.012 & 0.265 & 0.791\\
			38 & Home & -0.043 & 0.019 & -2.272 & \textbf{0.023}\\
			\motif{38} & Away & 0.054 & 0.018 & 2.971 & \textbf{0.003}\\
			46 & Home & 0.005 & 0.017 & 0.289 & 0.773\\
			\motif{46} & Away & -0.024 & 0.016 & -1.477 & 0.140\\
			74 & Home & 0.010 & 0.006 & 1.600 & 0.110\\
			\motif{74} & Away & -0.000 & 0.006 & -0.063 & 0.950\\
			78 & Home & -0.002 & 0.006 & -0.335 & 0.738\\
			\motif{78} & Away & 0.011 & 0.005 & 2.050 & \textbf{0.040}\\
			98 & Home & 0.046 & 0.036 & 1.267 & 0.205\\
			\motif{98} & Away & -0.033 & 0.040 & -0.836 & 0.403\\
			102 & Home & 0.015 & 0.014 & 1.092 & 0.275\\
			\motif{102} & Away & 0.025 & 0.013 & 1.944 & 0.052\\
			108 & Home & -0.008 & 0.019 & -0.442 & 0.658\\
			\motif{108} & Away & 0.027 & 0.019 & 1.418 & 0.156\\
			110 & Home & -0.032 & 0.007 & -4.351 & \textbf{0.000}\\
			\motif{110} & Away & 0.011 & 0.007 & 1.535 & 0.125\\
			238 & Home & -0.024 & 0.007 & -3.494 & \textbf{0.000}\\
			\motif{238} & Away & 0.038 & 0.007 & 5.789 & \textbf{0.000}\\
			\hline\hline
		\end{tabular}
	\end{table}
	
	Table~\ref{table:motif:ols:freq} shows the results of the OLS regression model. The F statistic for the model is 29.09, and the $p$-value is 0. It implies that game results have a correlation with the number of motifs. The 3-motifs encoded as 12, 38, and 238 for both home and away teams have significant correlations with the goal differential. The 3-motifs encoded as 12 are one-directional attack paths without passing back. The 3-motifs encoded as 38 are one-directional attack paths with two different passes. One possibility is that the original attack path $i\rightarrow j$ is blocked and changed to path $i\rightarrow k\rightarrow j$. Another possibility is to change the original two passes into a long pass to speed up the attack. These will lead to a reduction in the success rate of attacks. The 3-motif encoded as 238 is the strongest connected 3-motif. It suggests that there is no clear pass path among them, which may be due to the misplaced pass or the opponents' defense. The results of the model tell us that a team with more directed passes and efficient passes will have a higher goal differential.
	
	The 3-motifs encoded as 14, 110 for home teams, and 78 for away teams have significant correlations with the goal differential. The 3-motif encoded as 14 is similar to the 3-motif encoded as 12, which has a clear attack path among them. The 3-motifs encoded as 78 and 110 have a lot of bidirectional passes.

	\section{Discussion}
	
	In this paper, we investigate the passing activities in football games. We focus on the successful passing events (units of attacks) for both home and away teams. We find that home teams made fewer attacks than away teams in a game, which is different from the game results. It means that the success rate of attacks from home teams is higher than that from away teams. The length of each unit of attacks is close to a gamma fit, which may be due to the combination of randomizations of many attacks. 
	
	In order to find passing attributes from these random events, we use the social network analysis method to explore the characteristics of the passing network. We construct both home and away passing networks for each football game. The number of passes between two footballers is close to an exponential fit. Nearly half of the pass relationships only contain 1 or 2 passes. Thus, the original passing networks composed of all passes have high density and transitivity. To filter out the real cooperation from the passing events, we focus on the passing networks with a link weight over 2. Compared with away teams, network measures for home teams are slightly smaller.
	
	We investigate the statistics of 3-motifs for all the home and away teams. Our results show that 3-motifs with a large ratio of bidirectional links are more significant than the random network. It suggests that footballers tend to pass back in the process of attacks. Furthermore, we analyze the correlation between the results of games and 3-motifs. We find that teams with more directed passes and fewer passbacks will have a higher goal differential. Because a direct pass is usually effective for an offense, a pass back will extend the attack path and increase the possibility of attack failure.
	
	Our results for investigating passing networks show the 3-motifs can help us understand the micro-characteristics of players' passing behavior. Analyzing the frequency and distribution of these 3-motifs can help coaches and analysts understand the team's passing strategy, the interaction efficiency between players, and the possible direction of tactical optimization.

	\section*{Acknowledgments}
	
	This work was supported by the Fundamental Research Funds for the Central Universities.
	
	

\end{document}